\documentstyle[epsf]{article}
\makeatletter
\def\epsfile#1{\def\@psfile{}\parse@ps@parms{#1}}
\def\parse@ps@parms#1{%
  \@for\@epsfile:=#1\do{\expandafter\@setparms\@epsfile,}%
  \epsfbox{\@psfile}}
\def\@setparms#1=#2,{\@nameuse{@setps#1}{#2}}
\def\@setpsfile#1{\def\@psfile{#1}}
\def\@setpsheight#1{\epsfxsize=0pt\epsfysize=#1}
\def\@setpswidth#1{\epsfxsize=#1\epsfysize=0pt}
\def\@setpsscale#1{\def\epsfsize##1##2{#1##1}}
\makeatother
\begin{document}
\hfuzz=1pt
\setlength{\textheight}{8.5in}
\setlength{\topmargin}{0in}
\begin{center}
\LARGE {\bf Simulation and analysis of electron
cyclotron resonance discharges}
\\  \vspace{.75in}
\large {M. Ardehali}
\\ \vspace{.3in}
Research Laboratories,
NEC Corporation,\\
Sagamihara,
Kanagawa 229
Japan

\end{center}
\vspace{.20in}

\begin{abstract}
We describe in detail the method for 
Particle-in cell/Monte-Carlo simulation of electron 
cyclotron resonance (ECR) discharges. 
In the simulation,
electric and
magnetic fields
are obtained by solving Maxwell equations, and electrons
and ions are accelerated by solving equations of motion.
We consider two different cases:
(i) propagation of electromagnetic wave in the presence of 
a constant external magnetic field;
(ii) propagation of electromagnetic wave in the presence of 
a linearly decreasing magnetic field which corresponds to 
a realistic ECR discharge.
The simulation results indicate that at the resonance layer,
the electrons are heated by the
electromagnetic wave, and the
incoming wave amplitude is pronouncedly damped, with the 
wave hardly propagating through the ECR layer.
\end{abstract}
\pagebreak

In recent years, there has been increasing interest in high density
plasmas at low gas pressures for semiconductor wafer 
processing \cite{1}.
Unlike Reactive Ion Etching (RIE)
discharges in which the electrons mean free path is
of order of a few centimeters, in Electron Cyclotron Resonance (ECR)
discharges, the electrons
are confined by the external magnetic field and their mean free path
is of order of less than a millimeter.
Thus ECR discharges are 
capable of generating high density plasmas at low gas pressures and
low temperatures.
Because of these advantages, numerous experiments 
have been performed to study these discharges. However, the 
fundamental understanding of these discharges is not yet satisfactory.
The main goal of this work is to describe the method for
simulating ECR discharges using the
self consistent Particle-in-
cell/Monte-Carlo (PIC/MC) technique.

The present PIC/MC simulator uses
particle-in-cell (PIC)
scheme for charge-assignment-force-interpolation \cite {2}, and
Monte Carlo technique for collisions and scatterings.
In the simulation,
an electromagnetic wave with frequency
$\omega_0= 2.45$ GHz enters the system along the $z$ axis.
PIC/MC is used to model the interaction of the electromagnetic 
wave with 
the electrons and ions.

Very briefly, the
PIC/MC algorithm for ECR plasma
consists of the following five subroutines:

$(I)$ Interpolate the instantaneous velocities of MC particles
representing ions
and electrons to the grid points using PIC technique.
Once the velocity of ions and electrons at a grid is 
obtained,
the current density at the same grid can easily be calculated.

$(II)$ Solve the Maxwell's equation on a
spatially discretized mesh to obtain the electric and magnetic fields.
In the simulation, we assume that
the electric and magnetic fields do not vary along the
$x$ and $y$ axes, i.e., we only consider
$E_x(z,t)$, $E_y(z,t)$, $E_z(z,t)$, $B_x(z,t)$, $B_y(z,t)$,
$B_z(z,t)$. This assumption is justifiable since 
electrons are confined to a radius of less than a millimeter
by the external 
magnetic field. For example, for an electron temperature
of 2 eV, which corresponds to the electron velocity of $10^8$ cm/sec,
and  for a
magnetic field of $B=10^{-2}$ T,
the electron's radius is less than
$1$ millimeter (here we have used the formula
$r= \frac {\displaystyle m|v|}
{\displaystyle eB \sin \phi}$, here
$\phi$ is the angle between the momentary electron velocity vector
and the magnetic field vector, and $|v|$ is the 
absolute value of the electron velocity).

Assuming
the electric and magnetic fields do not have any 
variations along the $x$ and $y$ directions,
the Maxwell's equations
can be be written as
\begin{eqnarray} 
\frac{\delta E_x}{\delta t}=-c\frac{\delta B_y}{\delta z}-4\pi J_x \\
\frac{\delta B_y}{\delta t}=-c\frac{\delta E_x}{\delta z}, \\
\frac{\delta E_y}{\delta t}=c\frac{\delta B_x}{\delta z}-4\pi J_y, \\
\frac{\delta B_x}{\delta t}=c\frac{\delta E_y}{\delta z}, \\
\frac{\delta E_z}{\delta t}=-4\pi J_z, \\ 
\frac{\delta B_z}{\delta t}=0.
\end{eqnarray}
In the above equations, $E$'s, $B$'s  and $J$'s
represent the electric field and 
magnetic field and current density.
The boundary conditions are obtained 
by assuming that the circularly polarized waves
enter the plasma from the left and that
the transverse waves leave the system without
being reflected. Thus
the horizontal and vertical components of the electric and 
magnetic fields at $z=0$ and  $z=L$ are given by
\begin{eqnarray}
&&E_x(0)=E_0 \cos\omega t, \quad
B_x(0)=B_0 \sin\omega t,\quad
E_y(0)=E_0 \sin\omega t, \quad
B_y(0)=-B_0 \cos\omega t, \nonumber \\ 
&&E_x(L)=0, \qquad
B_x(L)=0, \qquad
E_y(L)=0, \qquad
B_y(L)=0, 
\end{eqnarray}
where $\omega$ is the frequency of the source, and
$E_0$ and $B_0$ are the electric and magnetic field of the source,
i.e., the electric and magnetic field at $z=0$.
The boundary condition of the longitudinal electric field is 
obtained by solving Guass Equations, i.e.,
\begin{eqnarray}
E_z(0)=4 \pi \sigma(0,t), \qquad
E_z(L)=4 \pi \sigma(L,t),
\end{eqnarray}
where $\sigma(0,t)$ and $\sigma(L,t)$
are the surface charge density at the
left and right boundaries which vary with time.
The surface charge density can be obtained from
\begin{eqnarray} \nonumber
\sigma(0,t)= \int\limits_{0}^{t} 
\left(J_{i} (0)-J_{e}(0) \right)dt', \\
\sigma(L,t)= \int \limits_{0}^{t} \left(J_{i} (L)-J_{e}(L) \right)dt'.
\end{eqnarray}
The longitudinal boundary conditions
imply that the total charge of the system including the boundaries is
zero.

The Maxwell's equation for the transverse wave is obtained by adding
and subtracting Eqs. $1$ and $2$ (or Eqs. $3$ and $4$), i.e.,
\begin{eqnarray}
\left(\frac{\delta }{\delta t} + c\frac{\delta }{\delta z} \right)
F_{x;y} =-4\pi J_{x;y}, \nonumber \\
\left( \frac{\delta }{\delta t} - c\frac{\delta }{\delta z} \right)
G_{x;y} =-4\pi J_{x;y}, 
\end{eqnarray}
where $F_{x}=E_x+B_y$, $G_{x}=E_x-B_y$,
$F_{y}=E_y+B_x$, and $G_{y}=E_y-B_x$. 
Note that the left-hand side of Eq. (10) can be considered as the
total derivative along the vacuum line $z=ct$. Thus if we 
assume $\Delta z= c\Delta t$, Eq. (10) may be discritized as \cite {3}
\begin{eqnarray}
\frac{F_{x;y}(t+\Delta t,z+c\Delta t)-F_{x;y}(t,z) }{\Delta t} = 
-4\pi J_{x;y}
\left(t+ \frac{\Delta t}{2}, z+c \frac{\Delta t}{2} \right),
\end{eqnarray}
or
\begin{eqnarray}
F_{x;y,j+1}^{n+1}=
F_{x;y,j}^{n}- 4 \pi \left [ \left(J_{x;y,j+1}+J_{x;y,j}\right)/2
\right] \Delta t.
\end{eqnarray}
Thus the summation of the $x$ component of the
electric field and the $y$ component of the magnetic field
at grid point $j+1$ and at the time $n+1$
depends on  the summation of the $x$ component of the
electric field and the $y$ component of the magnetic field
at grid point $j$ and at the time $n$
and on the average current density between grids $j$ and $j+1$
at time $t + \Delta t/2$.
Using the above technique, one can obtain
$E_x(z,t), E_y(z,t), B_x(z,t)$, and $B_y(z,t)$.

$(III)$ Interpolate the electric field and the magnetic field
from the grid points to the
location of particles. Once the electric and magnetic fields
at the location of particles are known, equations of motion 
can be solved.

$(IV)$ Integrate the equations of motion under the local and
instantaneous electric and magnetic fields. To move the 
particles, we have to solve Lorentz equation.
\begin{eqnarray}
\mbox{\boldmath $v$}^{n+1}=
\mbox{\boldmath $v$}^{n}+\frac{q}{m}\Delta t
\left[\mbox{\boldmath $E$}^{n}+
\frac{1}{2} \left(\mbox{\boldmath $v$}^{n+1}+
\mbox{\boldmath $v$}^{n} \right) \times 
\mbox{\boldmath $B$}^{n} \right]  -\Delta t \frac{g}{m}
\frac{\Delta \mbox{\boldmath $B_{ext}$}}{\Delta z},
\end{eqnarray}
where g is the magnetic moment and $\mbox{\boldmath $B_{ext}$}$ is the
external magnetic field. Note that
the simulation uses Leap-Frog technique, and hence
time $n$ refers to $t-\Delta t/2$ and
time $n+1$ refers to $t+\Delta t/2$.

Since $\mbox{\boldmath $v$}^{n+1}$ appears on both sides of the
above equation, one has to proceed very carefully.
To obtain the velocity at time $n+1$ from the velocity at
time $n$, we
use Boris's technique \cite {4}, which is
based on the following three steps:
\\
$(i)$ 
First we define velocity $\mbox{\boldmath $v$}^{-}$ as
\begin{eqnarray}
\mbox{\boldmath $v$}^{-} =
\mbox{\boldmath $v^n$}
+\frac{q \mbox{\boldmath $E$}}{m}
\frac{\Delta t}{2}
-\frac{g}{m}
\frac{\Delta \mbox{\boldmath $B_{ext}$}}{\Delta z} \frac{\Delta t}{2}
\end{eqnarray}
Next we define velocity
$\mbox{\boldmath $v'$}$ which is related to  
the velocity $\mbox{\boldmath $v$}^{-}$ by the following relation
\begin{eqnarray}
\mbox{\boldmath $v'$}=
\mbox{\boldmath $v$}^{-}+
\mbox{\boldmath $v$}^{-} \times 
\mbox{\boldmath $r$},
\end{eqnarray}
where the function $\mbox{\boldmath $r$}$ is
defined as 
$\mbox{\boldmath $r$}=
\displaystyle \frac{q \mbox{\boldmath $B$}}{m}  \frac{\Delta t}{2}$.
\\
$(iii)$
Finally we define velocity $\mbox{\boldmath $v^+$}$
\begin{eqnarray}
\mbox{\boldmath $v$}^{+}=
\mbox{\boldmath $v$}^{-}+
\mbox{\boldmath $v'$} \times 
\mbox{\boldmath $s$}
\end{eqnarray}
where the function \mbox{\boldmath $s$} is defined as
\begin{eqnarray}
\mbox{\boldmath $s$}=
\frac{2 \mbox{\boldmath $r$}}{1+ \mid r \mid^2}
\end{eqnarray}
Boris \cite {4} has shown that 
the velocity at time $t+ \Delta t$ can be obtained from the
following equation
\begin{eqnarray}
\mbox{\boldmath $v$}^{n+1}=
\mbox{\boldmath $v$}^{+} +\frac{q \mbox{\boldmath $E$}}{m}
\frac{\Delta t}{2} - \frac{g}{m}
\frac{\Delta \mbox{\boldmath $B_{ext}$}}{\Delta z} \frac{\Delta t}{2}
\end{eqnarray}
The equations of motions for 
electrons and ions at time $t$ is numerically integrated to 
obtain the position of the electrons and ions at time step 
$t+ \Delta t$.
\begin{eqnarray}
z(t + \Delta t)=z(t)+\mbox{\boldmath $v$} ^{n+1}
\Delta t
\end{eqnarray}
$(V)$ Use random numbers (Monte Carlos technique)
and collision cross sections to
account for scattering and ionizations.
The total electron-neutral
scattering cross section $\sigma_{total}(v)$ is
$\sigma_{total}(v)=\frac{\displaystyle K_{total}}{\displaystyle v}$,
where $v$ is the electron
velocity and $K_{total}=2 \times 10^{-8}$ $cm^3/s$ is the rate constant.
Ionizing collisions occur if the electron energy is larger than
a specific value (for example, for Argon the threshold energy is
$15$ eV).
An ionizing collision is modeled by loading a new electron
and ion at the position of the ionizing electron. The kinetic
energy after ionizing  collision is partitioned between the two
electrons.
Ion-ion charge exchange and ion-ion
elastic collisions are also included in the simulator.

In the simulator, the charged particles
move under the influence of the self-consistent electric
and magnetic fields and suffer collisions with neutral particles.
The neutral
gas density is chosen to be $2 \times 10^{-14} cm^{-3}$.
The size of the discharge is 24 cm and the number of
grids is $2667$.
Microwave with an amplitude of $0.16$ Gauss and 
at a frequency of $2.45$ GHz enters the 
system from the left along the $z$ axis 
and propagates through the discharge.

Figure 1 (a) [Fig. 1 (b)] shows the variations of
$E_x$, $E_y$ [$B_x$, $B_y$]
within the discharge in the absence of particles.
The incoming wave propagates through the discharge without 
attenuation.
These figures clearly demonstrate that the
subroutine solving
Maxwell equation is
working properly.

We now consider the coupling of
the electro-magnetic wave to the electrons and ions. 
First,
we briefly describe the fundamental
principles of the ECR discharges.
We consider an external 
magnetic field along the $z$ axis
with a magnitude of $B_{ext}$ (for simplicity we 
assume that the external electric field is zero). 
 An electron 
rotates around the magnetic field with a
frequency of $\omega_c=\frac{\displaystyle qB_{ext}}{\displaystyle mc}$
($\omega_c$ is $2.45$ GHz
when the external magnetic field is $B_{ext}=875$ Gauss).
We now assume that an electromagnetic wave with frequency
$\omega_0$ enters this system. If $\omega_0$ is much
smaller or much larger than $\omega_c$, the electron is not
heated by the incoming wave.
However, when $\omega_0=\omega_c$,
resonance condition is attained and the wave
energy is absorbed, leading to strong
acceleration of electrons.

In the simulation, 
we assume an electromagnetic wave with frequency 
$\omega_0= 2.45$ GHz enters the system from the left.
First we consider
two types of external magnetic fields: (1)
$B_{ext}=1875$ G which corresponds to an electron
cyclotron frequency of
$\omega_c=\frac{\displaystyle qB_{ext}}{\displaystyle mc} = 5.24 GHz$,
(2) $B_{ext}=875$ G which corresponds to an electron
cyclotron frequency of $2.45 GHz$.
Figure 2 (a) [Figure 2 (b)] shows the variation of the horizontal
component of the electric [magnetic] field
within the discharge. When $B_{ext}=875$ G, which
corresponds to electron cyclotron frequency of $2.45$ GHz and which
is equal to the frequency of the electromagnetic wave, resonance
occurs and the electric and magnetic fields of the incoming
electromagnetic wave are pronouncedly damped, with the wave
hardly propagating within the discharge.
In contrast, when 
$B_{ext}=1875$ G, which corresponds to an electron
cyclotron frequency of $5.25$ GHz
and which is much larger than the frequency of the incoming wave,
the electric and magnetic fields of the
electromagnetic wave propagate through the discharge
without attenuation (note that
we use CGS system where both electric and magnetic fields are
measured in Gauss).

Figure 3 shows the horizontal component of electron 
velocity within the discharge.
When 
$B_{ext}=875$ G, 
resonance
occurs and the electrons are heated by the
incoming electromagnetic wave. Thus the horizontal component of the
electron velocity increases sharply right at the boundary.
However,
When 
$B_{ext}=1875$ G, 
the electrons do not absorb much energy  from the incoming wave.
The electron velocity is therefore small at the boundary and
does not change rapidly within the discharge.

In the previous examples,
we assumed that the external magnetic field is
constant within the discharge. In an actual ECR discharge,
$B_{ext}$
drops along $z$ axis.
To model a realistic ECR discharge \cite {5}, we simulated
 a system where
the external magnetic field decreases linearly along the $z$ axis
so that at the center of the discharge,
the external magnetic field is $875$ G and hence resonance occurs.

Figs. 4 (a) and 4 (b) show
the trajectories of electrons 
at the center of discharge 
when the external magnetic fields of $1875$ G
and $875$ G. Although in the simulation, we trace the electrons only
along the $z$ direction, Figs. 4 (a) and 4 (b)
are obtained by integrating $V_x$ and $V_y$ over time.
Note that when 
$B_{ext}=875$ G,  i.e.,
when $\omega_c=\omega_0$, resonance condition is attained and 
electrons spiral around the external magnetic field.
In contrast, when 
$B_{ext}=1875$ G, i.e., 
when $\omega_c >> \omega_0$, 
resonance does not occur
and electrons do not spiral around
the magnetic field.

Figures $5 (a)$ 
and $5 (b)$ show the horizontal
and vertical
components of the
electric field within the discharge.
The incoming electromagnetic wave is entirely
absorbed by the electrons near the resonance layer, with 
both $E_x$ and $E_y$ dropping
rapidly near the ECR layer.
The incoming wave hardly propagates beyond the resonance layer.

Figures $6 (a)$ and $6 (b)$ show
the horizontal and vertical components of the electron velocity within
the discharge after $65$ cycles.
At the resonance layer, the
velocity of electrons
jumps significantly, indicating that the electrons 
absorb considerable energy
from the incoming electromagnetic wave.
However at other positions within the discharge,
the velocity of electrons
does not change rapidly, indicating that
the electrons absorb very little energy.
Of course after many cycles, 
the horizontal and vertical components of electron 
velocity become isotropic as the the electrons localized near
the resonance layer create new electrons by ionization as well
as diffusing
toward the boundaries. To clearly demonstrate the heating of the
electrons at the resonance layer, we present here the
simulation results after only 65 cycles.

In summary, PIC/MC technique has been used to investigate the
fundamental properties of an electropositive
ECR discharge. 
The simulation results indicate that at resonance layer, i.e., at
$\omega_c=\omega_0$, the incoming electromagnetic wave is
pronouncedly damped, 
leading to dramatic acceleration of electrons.
The simulation results also show that at resonance layer, electrons
spiral around the external magnetic field.
PIC/MC technique shows great promise for simulating more complex 
(for example electronegative)
discharges in two or three dimensions.

\pagebreak

\pagebreak
{\LARGE \bf Figure  Captions}
\\ 
{\bf IMPORTANT NOTE:}
\\
Figs. 1 (a) and 1 (b) are on the same page.
\\
Figs. 2 (a) and 2 (b) are on the same page.
\\
Figure 3 is on one page.
\\
Figs. 4 (a) and 4 (b) are on the same page.
\\
Figs. 5 (a) and 5 (b) are on the same page.
\\
Figs. 6 (a) and 6 (b) are on the same page.
\\
Fig. 1 (a) Profile of the  horizontal (solid line)
and vertical (dashed line) components of 
the electric field within the discharge in the absence of particles.
\\
Fig. 1 (b) Profile of the horizontal (solid line)
and vertical (dashed line) components of 
the magnetic field within the discharge in the absence of particles.
\\
Fig. 2 (a) Profile of the horizontal component of 
the electric field within the discharge when the external 
magnetic field is at 875 G (solid line) and at 1875 G (dashed line).
\\
Fig. 2 (a) Profile of the horizontal component of 
the magnetic field within the discharge when the external 
magnetic field is at 875 G (solid line) and at 1875 G (dashed line).
\\
Fig. 3 Profile of the horizontal component of 
electron velocity within the discharge when the external 
magnetic field is at 875 G (solid line) and at 1875 G (dashed line).
\\
Fig. 4 (a) Trajectories of electrons when the external magnetic field
is at 
$1875$ Gauss.
\\
Fig. 4 (a) Trajectories of electrons when the external magnetic field
is at $875$ Gauss.
\\
Fig. 5 (a) Profile of the horizontal component of 
electric field in an ECR discharge. Note that at the
center of the discharge, the external magnetic field
is $875$ G.
\\
Fig. 5 (b) Profile of the vertical component of 
electric field in an ECR discharge. 
\\
Fig. 6 (a) Profile of the horizontal component of 
electron velocity in an ECR discharge after $65$ cycles.
\\
Fig. 6 (b) Profile of the vertical component of 
electron velocity in an ECR discharge after $65$ cycles.

\pagebreak
\epsfile{file=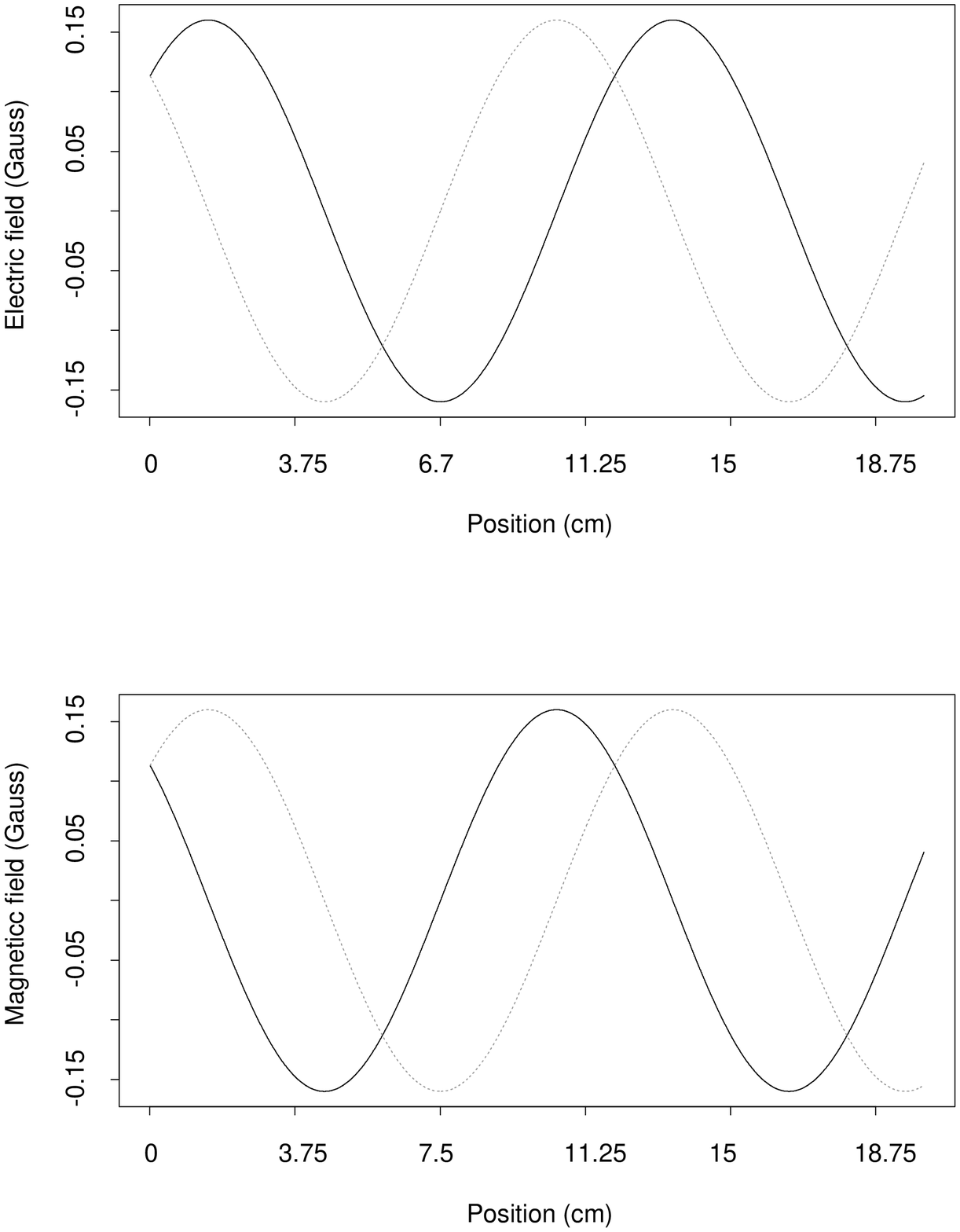,width=\textwidth}
\epsfile{file=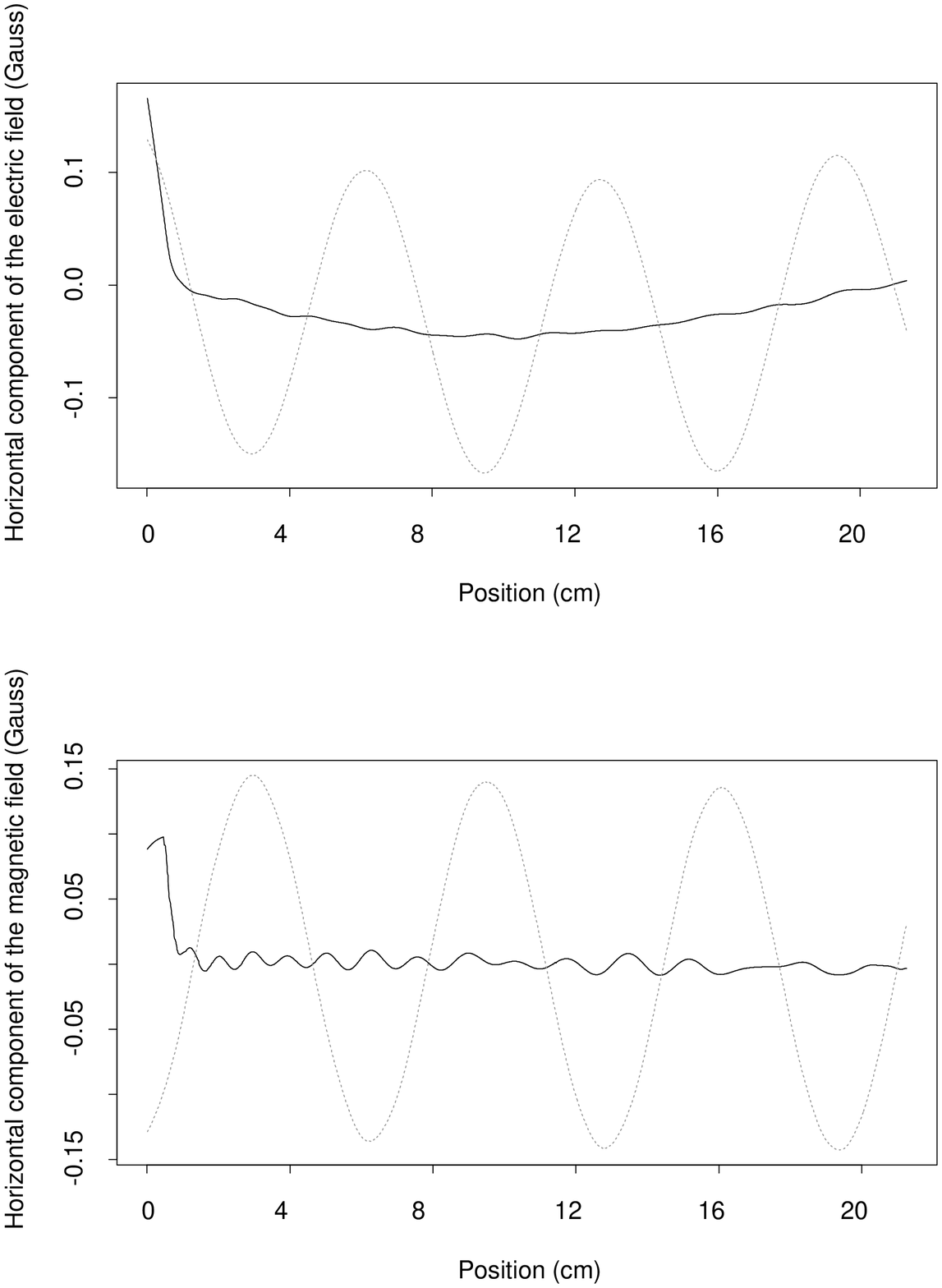,width=\textwidth}
\epsfile{file=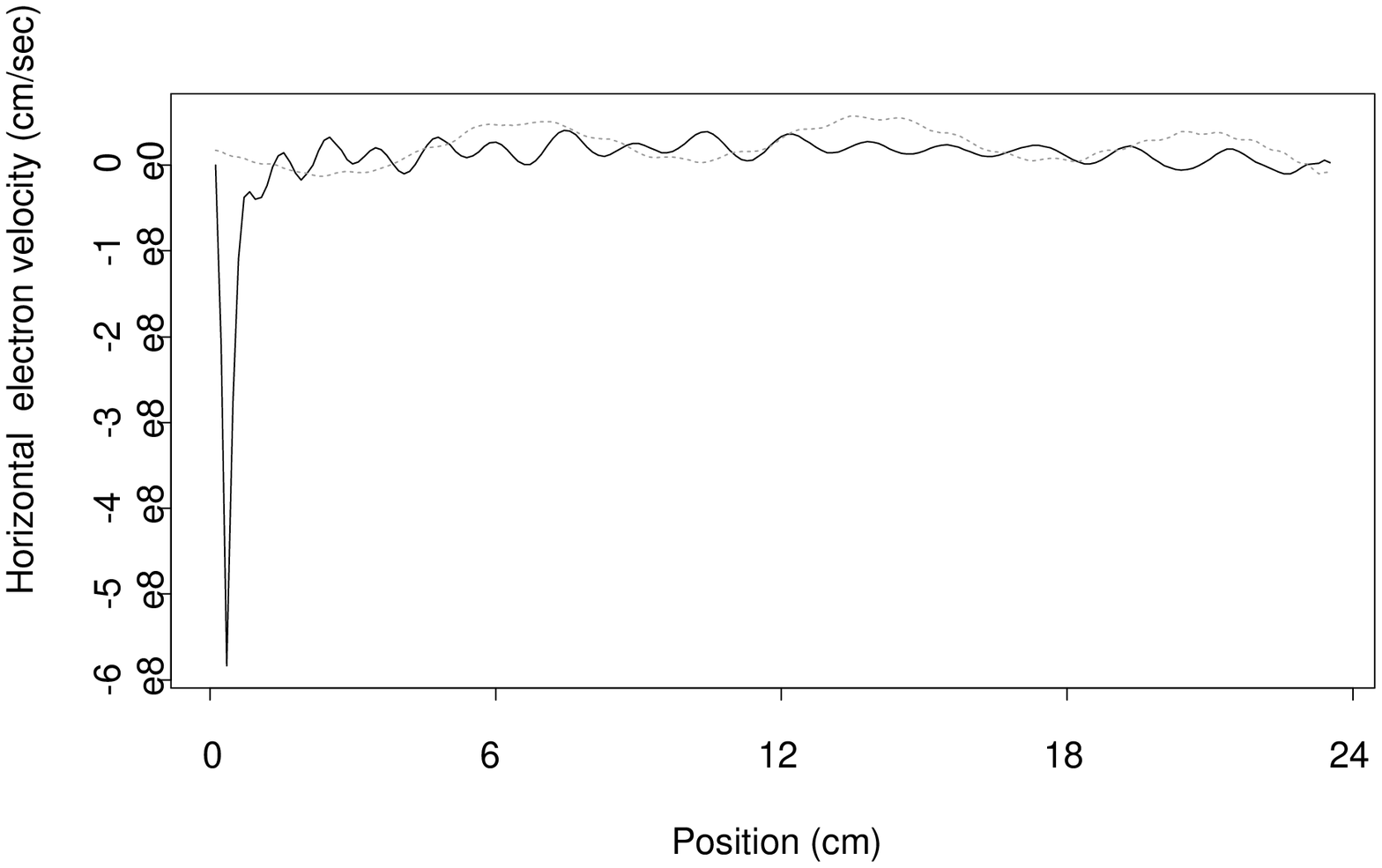,width=\textwidth}
\epsfile{file=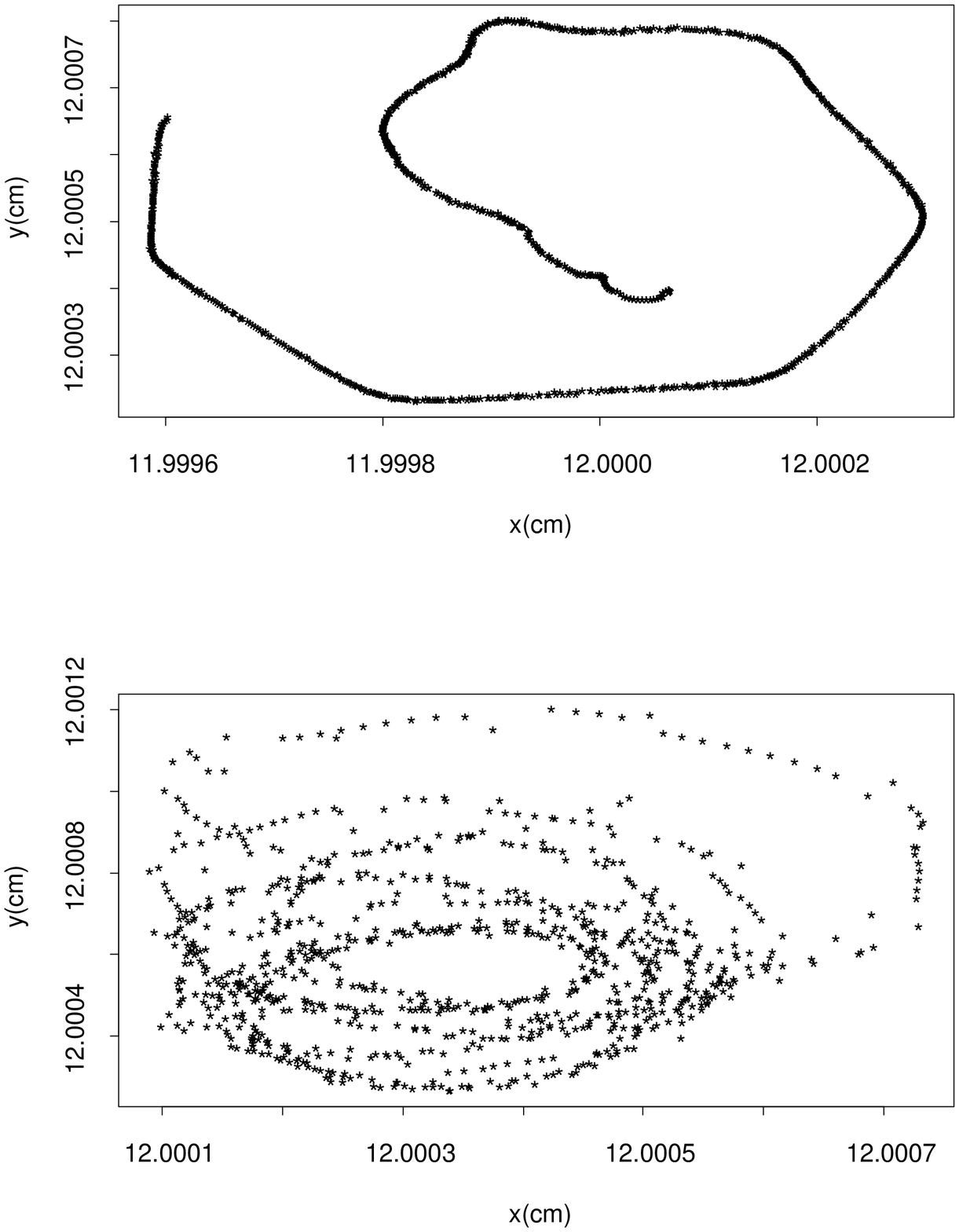,width=\textwidth}
\epsfile{file=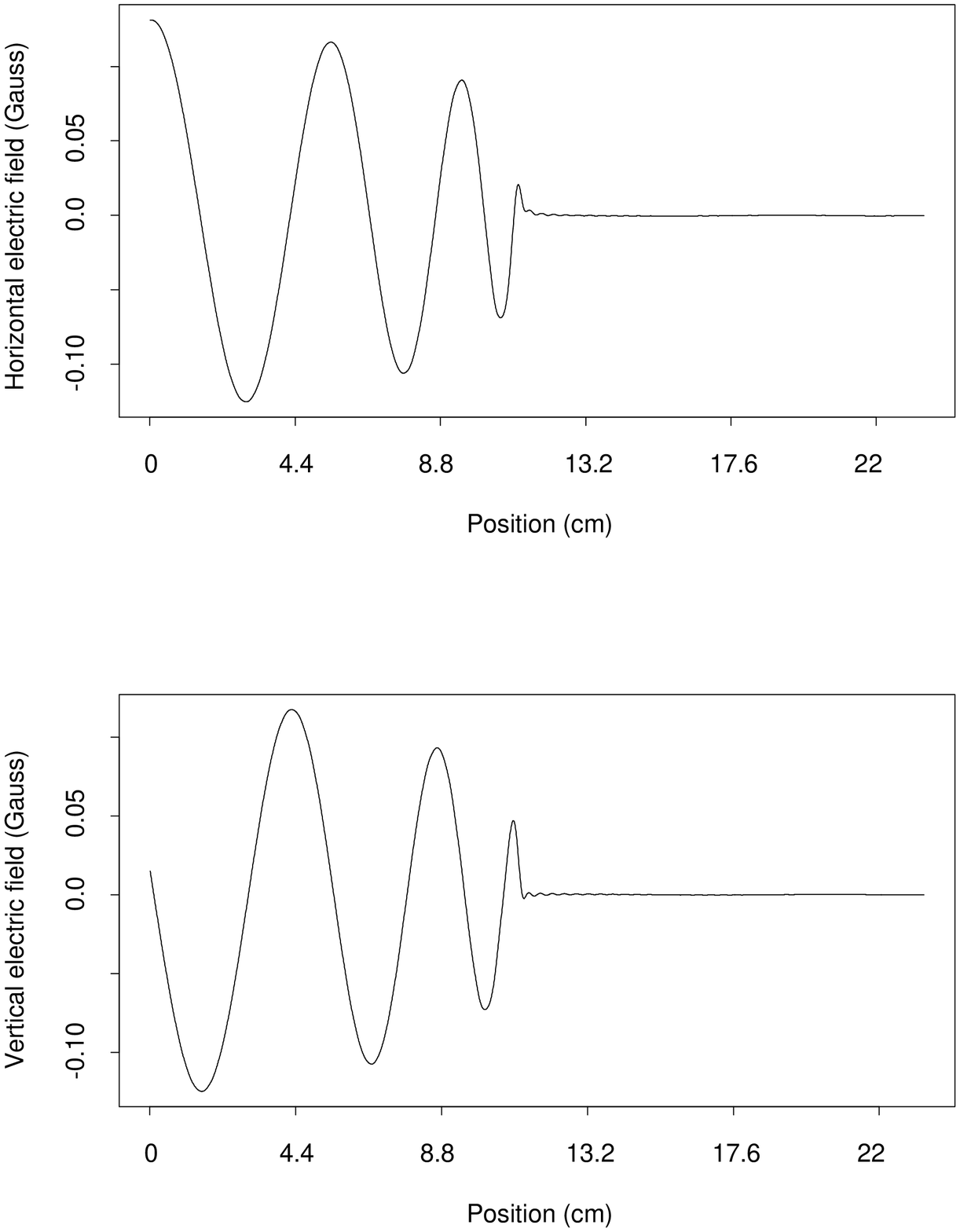,width=\textwidth}

\end{document}